\documentclass{jpsj-suppl}
\usepackage{times}
\usepackage{color}

\title{Magnetic properties of single crystalline CeMg$_{12}$}

\author{Pranab Kumar Das and A. Thamizhavel\thanks{E-mail address: thamizh@tifr.res.in} }

\inst{Department of Condensed Matter Physics and Materials Science, Tata Institute of Fundamental Research, Dr. Homi Bhabha Road, Colaba, Mumbai 400 005, India.}

\recdate{September 18, 2013}

\abst{Single crystal of CeMg$_{12}$ is obtained by Bridgman method. CeMg$_{12}$ crystallizes in the tetragonal structure with space group \textit{I4/mmm} (\#139). The Laue pattern confirms the tetragonal crystal structure of CeMg$_{12}$. We have studied the magnetic properties by measuring magnetic susceptibility, magnetization, heat capacity and electrical transport. Specific heat measurement shows that the compound orders at 1.2~K. We have measured the magnetic susceptibility in the temperature range 1.8 to 300 K, and the susceptibility data show that [100] crystallographic direction is the easy axis of magnetization. At high temperature inverse susceptibility varies linearly with temperature, and follows the Curie-Weiss behaviour. The effective moment is close to the free ion value thus indicating Ce is in trivalent state in this compound.}

\kword{CeMg$_{12}$, Bridgman Method, single crystal, crystal electric field, Curie-Weiss }

\begin{document}
\maketitle

\section{Introduction}

The 4$f$ electrons in the rare-earth based intermetallic compounds exhibit a wide range of magnetic properties and hence these compounds have been studied quite extensively.  Due to the vicinity of $f$ level and Fermi surface, the Ce-based intermetallic compounds are known to exhibit a variety of interesting properties apart from the usual trivalent behaviour, such as mixed valent, heavy fermion, Kondo lattice etc.  Most interestingly, the Ce-based intermetallic compound often exhibit quantum phase transition which is tuned by means of hydrostatic pressure, chemical doping or magnetic field.  These behaviors arise in Ce compounds due to the competing interaction between the Ruderman-Kittel-Kasuya-Yosida (RKKY) interaction and the Kondo effect.  A magnetic or non-magnetic ground state appears at low temperature depending on the strength of the exchange interaction $J_{cf}$ between the conduction electron ($c$) and $f$ electron. A magnetic ground state appears by the conduction electron mediated indirect exchange interaction between $4f$ electron moments, when the RKKY interaction is dominant.   Single crystals are often preferred to probe the magnetic properties due to their superior crystalline order.  Recently, we have been investigating the Ce-based binary compounds by growing them in single crystalline form~\cite{pkd1, pkd2}.  Like for example, we have grown the single crystal of the cubic compound CeMg$_3$ which is a heavy fermion antiferromagnet with a N\'{e}el temperature ($T_N$) of 2.6~K.  The electrical resistivity of CeMg$_3$ exhibited $-ln$(T) behaviour and the magnetization measurement revealed a reduced magnetic moment thus suggesting CeMg$_3$ as a heavy fermion Kondo lattice system.  The crystal field calculations on CeMg$_3$ revealed that the ground state is a $\Gamma_7$ doublet and the excited quartet state was separated by about 190~K.  An estimate of Kondo temperature ($T_{\rm K}$) was made from the heat capacity data of CeMg$_3$ and it was found to be 3.6~K which is of the same order as that of the $T_{\rm N}$.  In continuation to our studies on the Ce-Mg compounds, we report here on the crystal growth and anisotropic magnetic properties of magnesium rich CeMg$_{\rm 12}$ compound.

\section{Experiment}

The binary phase diagram of Ce and Mg has been investigated by Nayeb-hasehmi and Clark~\cite{Nayeb} and reported six different stable compounds of Ce and Mg, of which CeMg$_3$ and CeMg$_{12}$ melt congruently. Due to the high vapor pressure of Mg, we have adopted Bridgman method in protected molybdenum crucible to grow the single crystal of CeMg$_{12}$. High pure metals of Ce (99.9\%) and Mg (99.99\%) were taken in the stoichiometric ratio of 1:12, with a little excess of Mg, in a sharp tipped alumina crucible and sealed in a molybdenum tube in argon atmosphere. The molybdenum tube was then subsequently sealed in a quartz ampoule, in a vacuum of 10$^{-6}$~Torr, to prevent the oxidation of molybdenum crucible at high temperature.  The sealed quartz ampoule was loaded into a box type furnace.  The temperature of the furnace was then raised to 800~$^\circ$C and held at this temperature for 24 hours.  Then the furnace was cooled down rapidly down to 630~$^{\circ}$C, followed by a very slow cooling rate of 1~$^{\circ}$C/hr down to 580~$^{\circ}$C, finally the furnace was switched off to cool down to room temperature.  Bulk shiny single crystals were obtained by gently taping the alumina crucible. The phase purity of the sample was checked by means of powder x-ray diffraction (XRD) using monochromatic Cu-K$_{\alpha}$ radiation with wavelength 1.5406~\AA.  The XRD pattern showed that the sample is phase pure and possesses the space group \textit{I4/mmm}.  There were few impurity peaks corresponding to some un-reacted Mg. The lattice constants were estimated to be $a~=~10.332$~\AA~ and $c~=~5.961$~\AA~  which matches well with the previous structural report~\cite{Johnson}. The unit cell volume is 636.35~\AA$^3$ and the nearest Ce~$-$~Ce  distance is 5.961~\AA. The crystals were oriented along the two principal crystallographic directions viz., along [100] and [001] directions by means of Laue diffraction.  Well defined Laue diffraction spots together with the four fold symmetry confirmed the tetragonal crystal structure and the good quality of the sample.  We have also confirmed the stoichiometry of the grown single crystals by performing an energy dispersive analysis by x-ray (EDAX).    

\section{Results and Discussion}

\subsection{Heat Capacity}

\begin{figure}
\begin{center}
\includegraphics[width=0.5\textwidth]{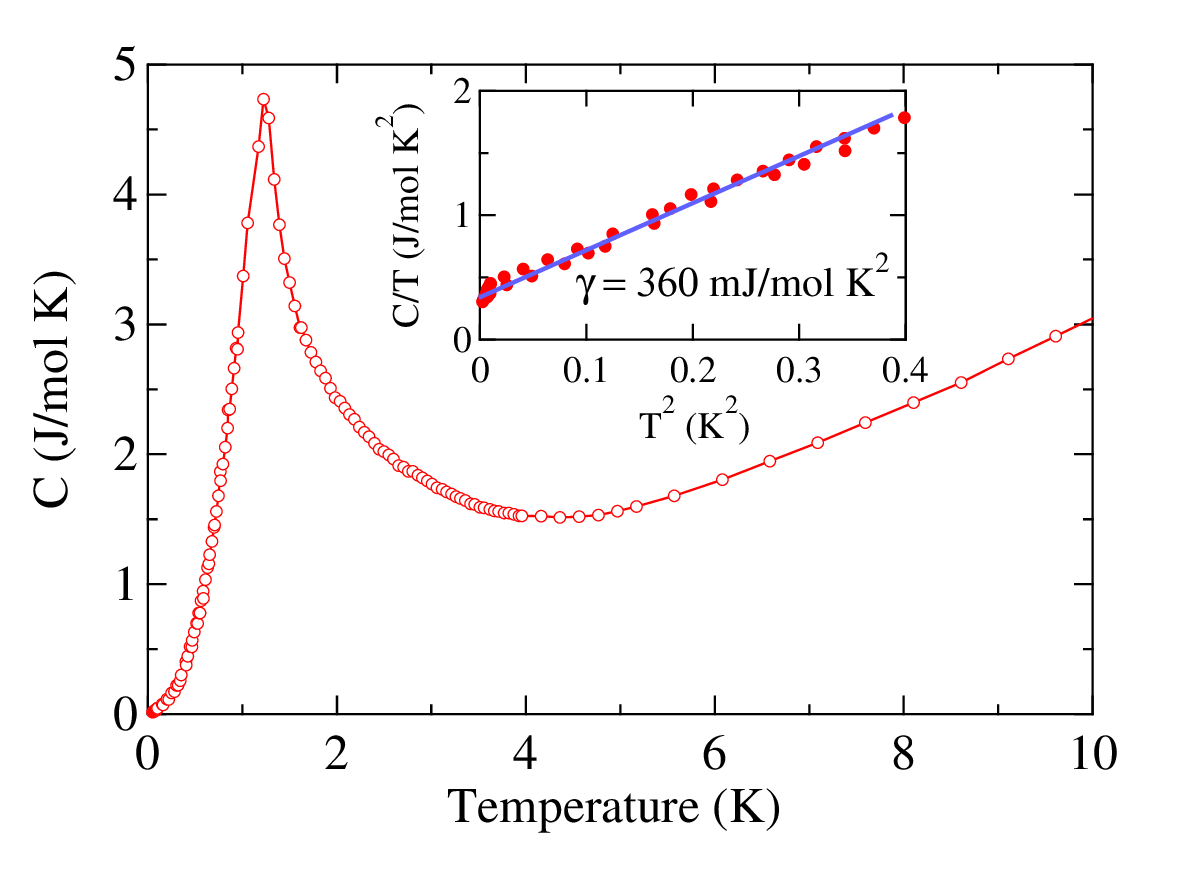}
\end{center}
\caption{(Color online) Temperature dependence of heat capacity of CeMg$_{12}$.  The inset shows the low temperature part of the $C/T$ versus $T^2$ plot.}
\label{fig4}
\end{figure}

The temperature dependence of specific heat capacity measured in a Quantum Design physical property measurement system (PPMS), in the temperature range from 0.05 to 10~K is shown in Fig.~\ref{fig4}. It is evident from the figure that a huge jump in the specific heat is observed at 1.2~K suggesting a bulk magnetic ordering $T_{\rm ord}$ in CeMg$_{12}$.  The jump in the specific heat is broad thus signalling a short range magnetic order above the $T_{\rm ord}$. The inset shows the low temperature part of the $C/T$ versus $T^2$ plot.    An estimation of the Sommerfeld coefficient $\gamma$ was made by the linear extrapolation of the $C/T$ versus $T^2$ plot.  The obtained $\gamma$ value is 360~mJ/K$^2$~mol. Although the $\gamma$ value is large, since the magnetic ordering temperature is at 1.2~K and the crystal field split first excited state (to be discussed later) is at 30~K, one cannot claim this compound to be a heavy fermion compound. We could not calculate the magnetic entropy as we did not have the non-magnetic analogue of CeMg$_{12}$ to subtract the phonon contribution.  

\subsection{Electrical Resistivity}
\begin{figure}
\begin{center}
\includegraphics[width=0.5\textwidth]{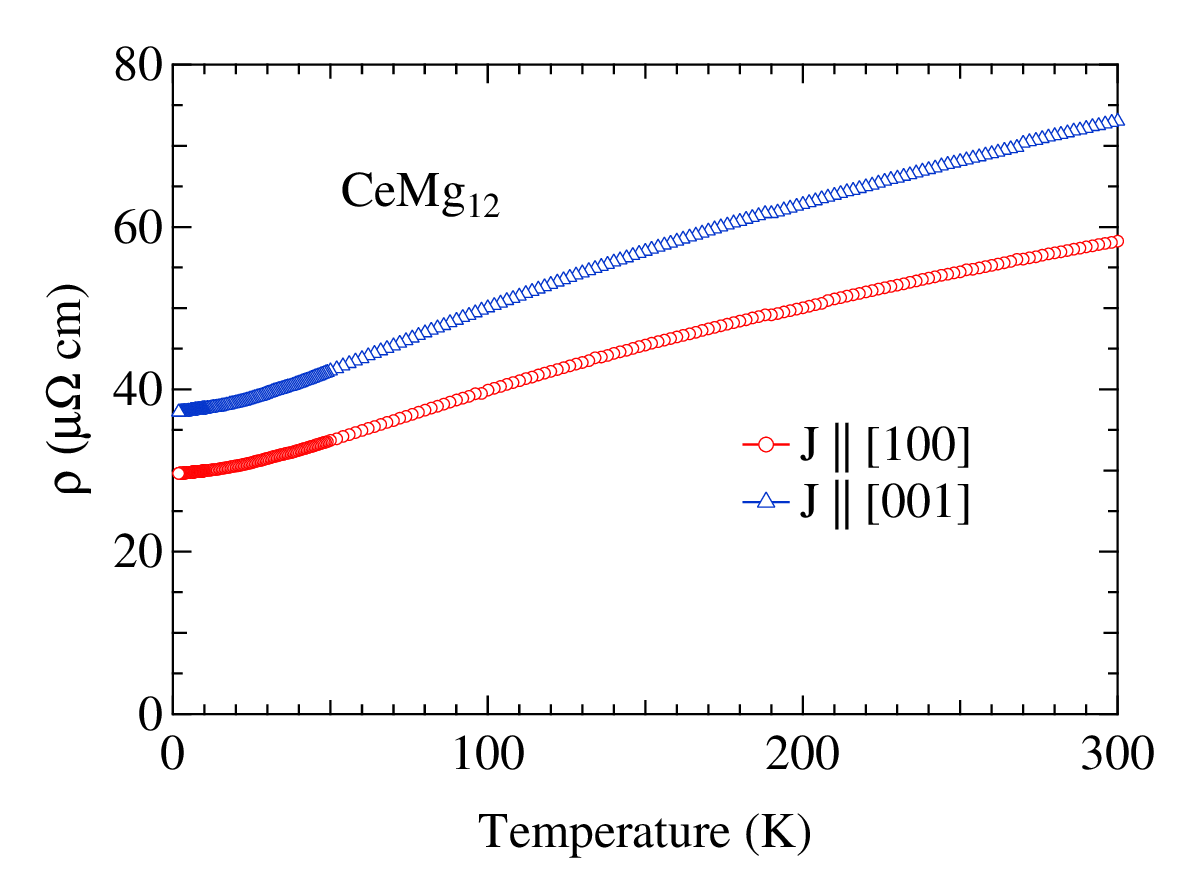}
\end{center}
\caption{(Color online) Temperature dependence of electrical resistivity along the two principal crystallographic directions.}
\label{fig3}
\end{figure}

The temperature dependence of electrical resistivity, measured in the temperature range from 1.8 to 300~K is shown in Fig.~\ref{fig3}.  As it is evidenced from the figure, there is a large anisotropy in the resistivity when the current is passed along the [100] and [001] direction.  Since the resistivity was measured down only to 1.8~K, the magnetic ordering observed at 1.2~K is not seen in the resistivity.  The electrical resistivity decreases with decreasing temperature showing a metallic behaviour.  A broad hump centered around 100~K is seen in the electrical resistivity, indicating the thermal population of the excited levels of the $2J+1$ crystal field levels. Our crystal field calculation, to be discussed later, confirms this.  No signature of Kondo effect is observed at low temperature.   

\subsection{Magnetic susceptibility and magnetization}
\begin{figure}
\begin{center}
\includegraphics[width=0.8\textwidth]{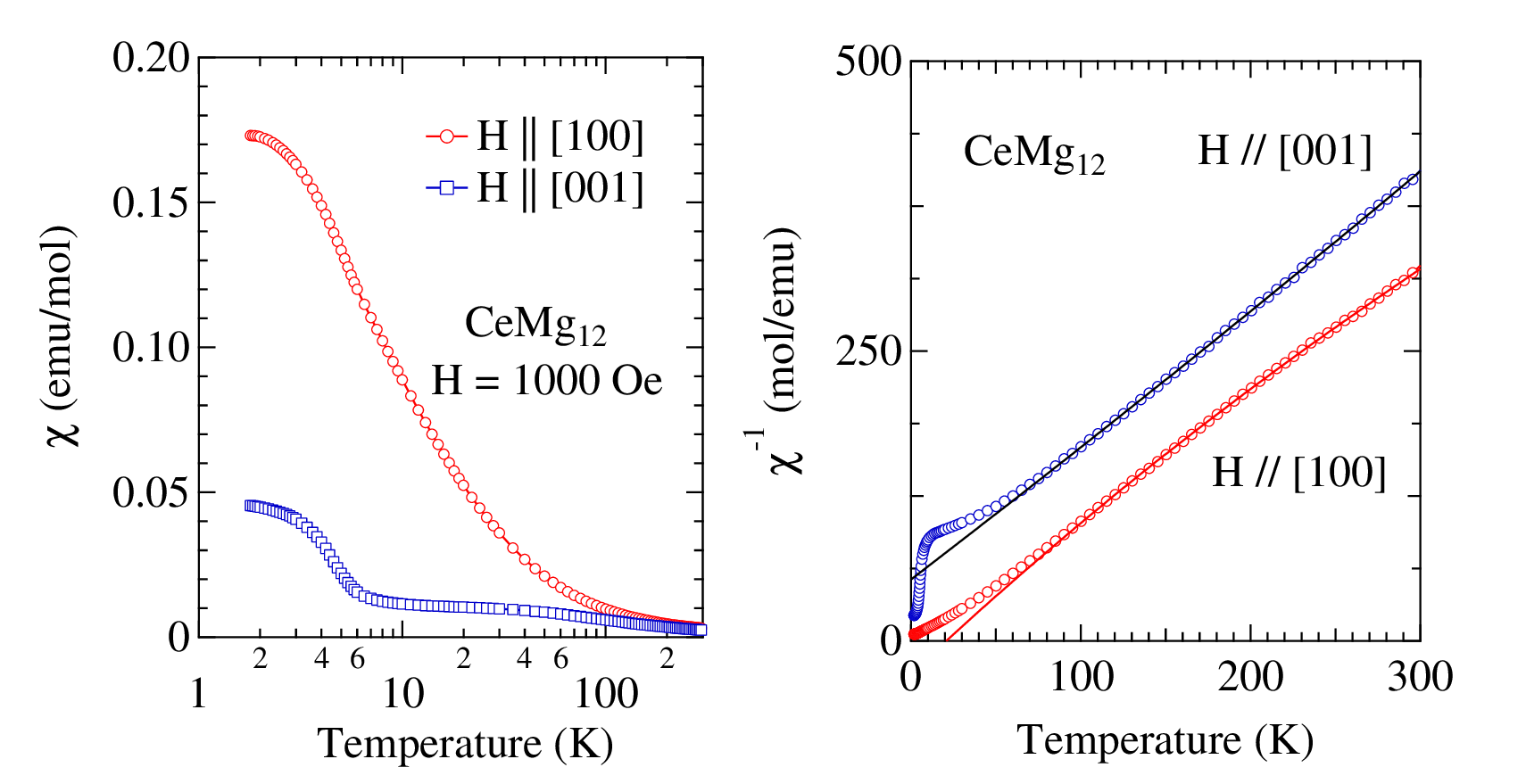}
\end{center}
\caption{(Color online) (a) Temperature dependence of magnetic susceptibility of CeMg$_{12}$ along the two principal crystallographic directions, and (b) Inverse magnetic susceptibility of CeMg$_{12}$.  The solid lines are fit to the Curie-Weiss law.}
\label{fig1}
\end{figure}

The temperature dependence of magnetic susceptibility of CeMg$_{12}$ measured in the range from 1.8 to 300~K in an applied magnetic field of 1 kOe along the two principal crystallographic directions is shown in Fig.~\ref{fig1}(a).  There is a large anisotropy in the magnetic susceptibility reflecting the tetragonal crystal structure. The magnetic susceptibility along the [100] direction is larger than along the [001] direction, thus indicating that [100] direction is the easy axis of magnetization.  Figure~\ref{fig1}(b) shows the inverse susceptibility plot. At high temperature, in the range 100 to 300~K the susceptibility obeys the modified Curie-Weiss law $\chi = \chi_0 + C/(T-\theta_p)$, where $\chi_0$ is the temperature-independent part of the magnetic susceptibility and $C$ is the Curie constant.  The main contributions to $\chi_0$ include the core-electron diamagnetism and the susceptibility of the conduction electrons. From the modified Curie-Weiss fitting we obtained $\chi_0$ = 4.561~$\times$~10$^{-4}$ emu/mol, $\theta_{\rm p}$~=~20.86~K and $\mu_{\rm eff}$~=~2.44~$\mu_{\rm B}$/Ce for $H~\parallel$~[100] and $\chi_0$ = -1.289~$\times$~10$^{-4}$ emu/mol, $\theta_{\rm p}$~=~$-$47.5~K and $\mu_{\rm eff}$~=~2.68~$\mu_{\rm B}$/Ce for $H~\parallel$~[001] direction. The deviation from the linear behaviour in the inverse magnetic susceptibility below 100~K is mainly attributed to the crystal electric field effect. 

\begin{figure}
\begin{center}
\includegraphics[width=0.5\textwidth]{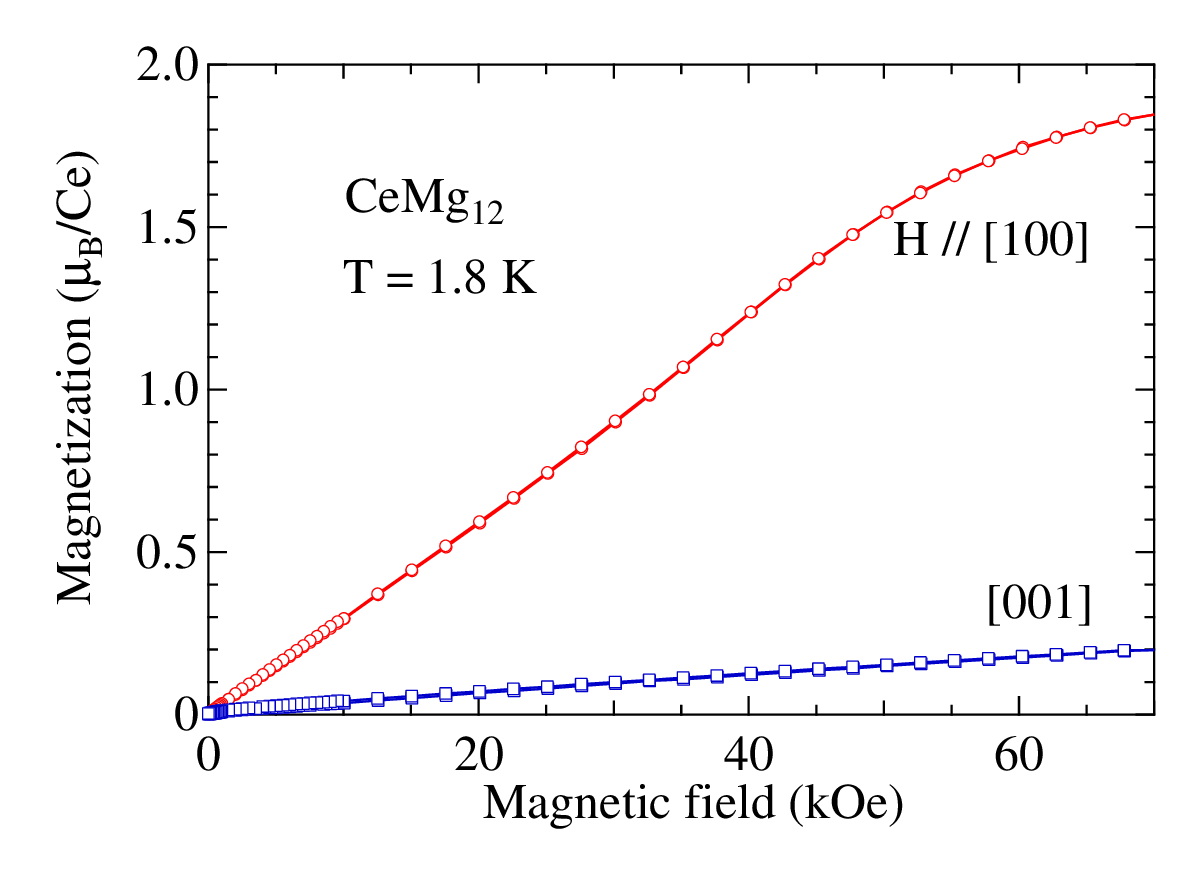}
\end{center}
\caption{(Color online) Isothermal magnetization measured at $T$~=~~1.8~K of CeMg$_{12}$ along the two principal crystallographic directions.}
\label{fig2}
\end{figure}

The isothermal magnetization measured at 1.8~K is shown in Fig.~\ref{fig2}. The magnetization for $H~\parallel~$ [100] increases more rapidly than along the [001] direction. For fields greater than 35~kOe, the magnetization shows some spin re-orientation, in spite of the fact that the magnetization measurement was done at 1.8~K, which is 0.6~K above the $T_{\rm ord}$~=~1.2~K. This confirms some sort of short range magnetic ordering above $T_{\rm ord}$. Furthermore, the magnetization shows a saturation behaviour for fields around 7~T. The magnetization reaches a value of 1.85~$\mu_{\rm B}$/Ce, which is close to the free ion value of Ce$^{3+}$ ($g_J J = \frac{6}{7} \times \frac{5}{2} = 2.14)$.  The magnetization for $H~\parallel$~[001] increases linearly and reaches only 0.2~$\mu_{\rm B}$/Ce at 70~kOe, indicating that [001] is the hard axis of magnetization.

We have performed the crystal field analysis on the magnetic susceptibility data to estimate the crystal field split energy levels of the degenerate $(2J+1)$ levels.  The Ce atom in CeMg$_{12}$ occupies the $2a$ Wyckoff's position in the space group $I4/mmm$ and hence possesses the tetragonal site symmetry $4/mmm$.  For the tetragonal site symmetry, the sixfold-degenerate levels of the $J = 5/2$ multiplet will split into three doublets.  The crystal electric field Hamiltonian for the Ce-atom in tetragonal site symmetry is given by, 
\begin{equation}
\label{eqn1}
\mathcal{H}_{{\rm CEF}}=B_{2}^{0}O_{2}^{0}+B_{4}^{0}O_{4}^{0}+B_{4}^{4}O_{4}^{4},\end{equation}
where $B_l^m$ and $O_l^m$ are the CEF parameters and the Stevens operators, respectively~\cite{Hutchings,Stevens}.  The magnetic susceptibility including the molecular field contribution $\lambda_{i}$, which represents the exchange interaction among Ce magnetic moments, is given
by
\begin{equation}
\label{eqn5}
\chi^{-1}_{i} = \chi_{{\rm CEF}i}^{-1} - \lambda_{i}.
\end{equation}

The expression for the magnetic susceptibility based on the CEF model is given by the following expression~\cite{pkd2}.

\begin{equation}
\label{eqn3}
\chi_{{\rm CEF}i} = N(g_{J}\mu_{\rm B})^2 \frac{1}{Z}
\left(\sum_{m \neq n} \mid \langle m \mid J_{i} \mid n \rangle
\mid^{2} \frac{1-e^{-\beta \Delta_{m,n}}}{\Delta_{m,n}}e^{-\beta
E_{n}} +  \sum_{n} \mid \langle n \mid J_{i} \mid n \rangle
\mid^{2} \beta e^{-\beta E_{n}} \right),
\end{equation}

where $g_{J}$ is the Land\'{e} $g$\,-\,factor, $E_{n}$ and $\mid\!n \rangle$ are the $n$th eigenvalue and eigenfunction, respectively.  $J_{i}$ ($i$\,=\,$x$, $y$ and $z$) is a component of the angular momentum,  and
$\Delta_{m,n}\,=\,E_{n}\, - \,E_{m}$, $Z\,=\,\sum_{n}e^{-\beta
E_{n}}$ and $\beta\,=\,1/k_{\rm B}T$.  For calculating the magnetization we have used the following Hamiltonian which includes the Zeeman term and molecular field term,

\begin{equation}
\label{eqn4}
\mathcal{H} = \mathcal{H_{\rm CEF}} - g_{\rm J} \mu_{\rm B} J_i (H_i + \lambda_i M_i),
\end{equation}
where,
\begin{equation}
\label{eqn6}
M_i = g_{\rm J} \mu_{\rm B} \sum_{n} \mid \langle n \mid J_{i} \mid n \rangle \frac{e^{-E_n /k_{\rm B}T}}{Z} ~~ (i = x, y, {\rm and}~z).
\end{equation}

\begin{figure}[h]
\begin{center}
\includegraphics[width=0.5\textwidth]{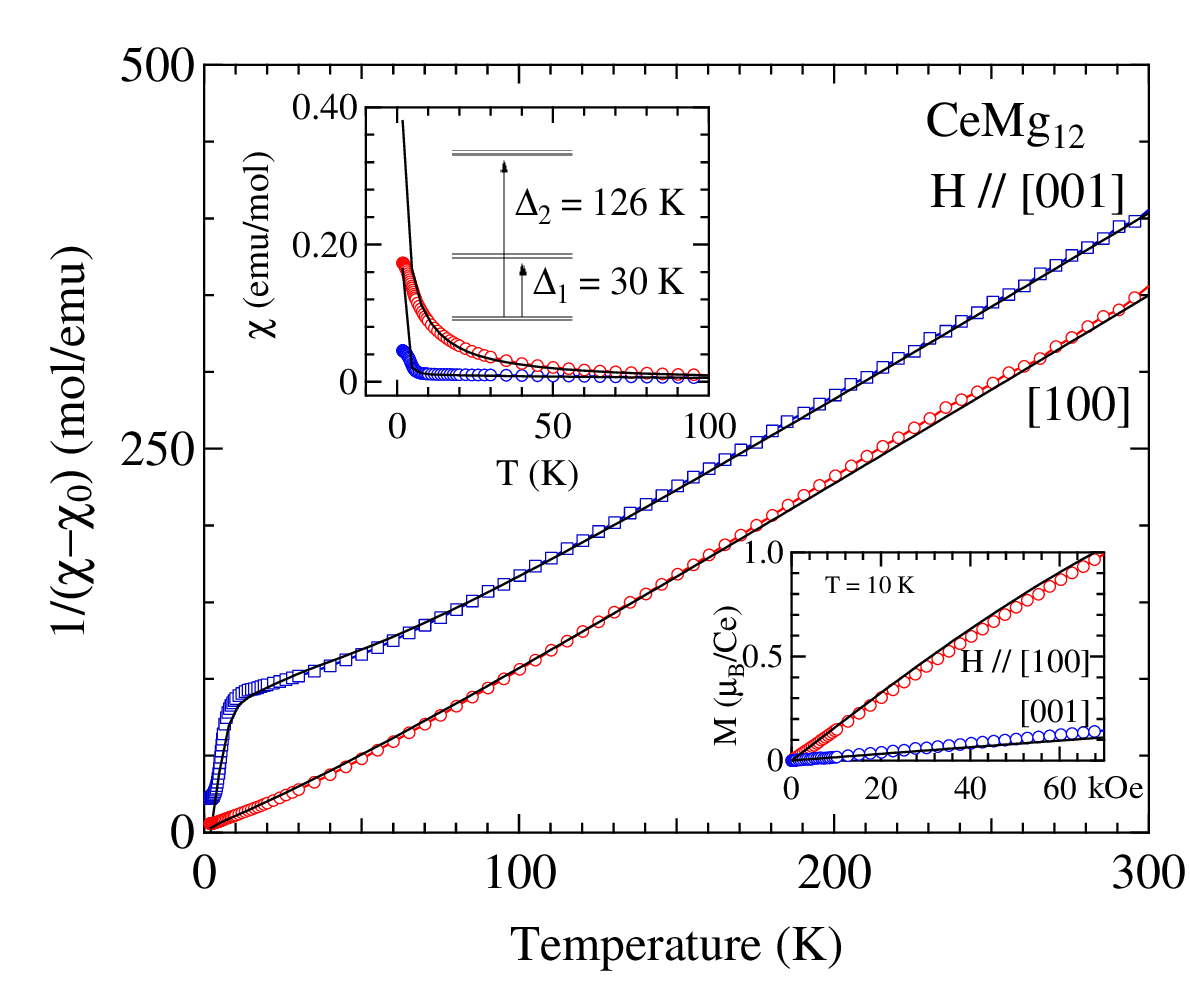}
\end{center}
\caption{(Color online) Inverse magnetic susceptibility of CeMg$_{12}$, the black solid lines are the calculated curves based on the crystal electric field calculations.  The obtained energy levels are shown in the inset of the figure.  The calculated CEF susceptibility is shown in the top inset.  The bottom inset shows the magnetization data measured at $T~=~10$~K, and the solid lines are the calculated CEF magnetization curves along the two principal crystallographic directions. }
\label{fig6}
\end{figure}

For the purpose of crystal field analysis, we have plotted the experimental results on susceptibility in the form of $1/(\chi - \chi_0)$. An estimate of the crystal field energy levels are obtained by diagonalizing the Hamiltonian.   The crystal field parameters thus obtained corresponds to $B_2^0$~=~6.8~K, $B_4^0$~=~0.022~K and $B_4^4$~=~ $-0.75$~K and the molecular exchange field constant $\lambda_{\rm 100}$~=~ 0 and $\lambda_{\rm 001}$~=~30~mol/emu. The positive value of the $\lambda$ along the [001] direction is suggesting a ferromagnetic ordering in CeMg$_{12}$, the calculated susceptibility diverges at low temperature along this direction also supports the evidence of ferromagnetic ordering. However, further low temperature measurements are necessary to confirm the nature of magnetic ordering in CeMg$_{\rm 12}$.  The  difference in the molecular field contributions for [100] and [001] directions, clearly indicate the anisotropic magnetic interaction between the Ce moments.  The sign of the $B_2^0$ parameter in general determines if the system possesses an easy plane or easy axis anisotropy~\cite{Loidl}.  Since $B_2^0$ is positive in CeMg$_{12}$  it has an easy plane magnetization. The obtained energy levels are $\Delta_{1}$~=~30~K and $\Delta_{2}$~=~126~K.  A detailed neutron diffraction study and the high temperature heat capacity measurements are necessary to clarify this estimated crystal field split level schemes.

\section{Summary and Conclusion}

We have successfully grown the single crystal of CeMg$_{12}$ by Bridgman method.  The grown crystals were oriented along the crystallographic directions and their anisotropic physical properties have been investigated.  Our heat capacity measurement shows that CeMg$_{\rm 12}$ orders magnetically at 1.2~K.  From the specific heat capacity and magnetization measurements we speculate that there may be a short-range magnetic ordering. This remains to be investigated by microscopic measurements like neutron diffraction experiment. There is a large anisotropy in the magnetic susceptibility and the electrical resistivity. The Ce-atom in CeMg$_{12}$ exhibit usual trivalent behaviour. The magnetization value reaches 1.8~$\mu_{\rm B}$/Ce in a field of 70~kOe. The crystal field analysis of the magnetic susceptibility data revealed that the $2J+1$ degenerate levels split into three doublets and the splitting energies were estimated to be  30~K and 126~K, corresponding to the first and the second excited states.


\begin{thebibliography}{9}

\bibitem{pkd1}Pranab Kumar Das, Neeraj Kumar, R. Kularni, S. K. Dhar and A. Thamizhavel, J. Phys. Condens. Matter {\bf 24} 146003 (2012).

\bibitem{pkd2}Pranab Kumar Das, Neeraj Kumar, R. Kulkarni and A. Thamizhavel, Phys. Rev. B {\bf 83}, 134416 (2011). 

\bibitem{Nayeb}B. Nayeb-hashemi and J. Clark, Bull. Alloy Phase Diagrams {\bf 9}, 916 (1988).

\bibitem{Johnson}Q. C. Johnson, Gordon S. Smith, D. H. Wood, E. M. Cramer, Nature {\bf 204}, 600 (1964).

\bibitem{Hutchings} Hutchings M T 1965 \textit{Solid State Physics: Advances in Research and Applications} \textbf{Vol.16} ed F Seitz and B Turnbull (New York: Academic) p.227.

\bibitem{Stevens} Stevens K W H 1952 \textit{Proc. Phys. Soc.} (London)  \textbf{Sect.A65} 209.

\bibitem{Loidl}A. Loidl, K. Knorr, G. Knopp, A. Krimmel, R. Caspary, A. B\"{o}hm, G. Sparn, C. Geibel, F. Steglich and A. P. Murani, 1992 \textit{Phys. Rev. B} {\bf 46}, 9341.


\end{thebibliography}
\end{document}